# Multi-energy X-ray linear-array detector enabled by the side-illuminated metal halide scintillator

*Peng Ran, Qingrui Yao, Juan Hui, Yirong Su, Lurong Yang, Cuifang Kuang, Xu Liu, Yang (Michael) Yang\**

P. Ran, Q. Yao, J. Hui, Y. Su, L. Yang, C. Kuang, X. Liu, Y. Yang
State Key Laboratory of Modern Optical Instrumentation, Institute for Advanced Photonics, College of Optical Science and Engineering, Zhejiang University, Hangzhou, 310063, China.
E-mail: yangyang15@zju.edu.cn

Conventional scintillator-based X-ray imaging typically captures the full spectral of X-ray photons without distinguishing their energy. However, the absence of X-ray spectral information often results in insufficient image contrast, particularly for substances possessing similar atomic numbers and densities. In this study, we present an innovative multi-energy X-ray linear-array detector that leverages side-illuminated X-ray scintillation using emerging metal halide $Cs_3Cu_2I_5$. The negligible self-absorption characteristic not only improves the scintillation output but is also beneficial for improving the energy resolution for the side-illuminated scintillation scenarios. By exploiting Beer's law, which governs the absorption of X-ray photons with different energies, the incident X-ray spectral can be reconstructed by analyzing the distribution of scintillation intensity when the scintillator is illuminated from the side. The relative error between the reconstructed and measured X-ray spectral was less than 5.63 %. Our method offers an additional energy-resolving capability for X-ray linear-array detectors commonly used in computed tomography (CT) imaging setups, surpassing the capabilities of conventional energy-integration approaches, all without requiring extra hardware components. A proof-of-concept multi-energy CT imaging system featuring eight energy channels was successfully implemented. This study presents a simple and efficient strategy for achieving multi-energy X-ray detection and CT imaging based on emerging metal halides.

## 1. Introduction

Scintillator-based X-ray detection and imaging play a crucial role in diverse fields such as medical diagnosis and industrial inspection[1]. In a conventional X-ray imaging system, X-rays are directed perpendicularly onto the scintillator detection plane. Upon passing through the test object, the remaining X-ray photons of all energies are absorbed by the scintillator and converted into visible light. The resulting X-ray images are typically displayed in black-and-white, as the image contrast is determined by the total integration of X-ray photons across all energy ranges[2]. However, this energy-integration approach often leads to low contrast for objects with similar atomic numbers and densities[3], the development of X-ray energy-resolving techniques has revolutionized this scenario, enabling significantly improved material discriminability[4]. Although energy-resolving photon-counting detectors in direct conversion mode have started to become available in high-end computed tomography (CT) systems, they still encounter certain limitations[5]. These limitations include the high cost associated with growing spectroscopic-grade crystals[6], performance issues under high-flux X-ray conditions due to "pile-up" effects[7], and limited availability of pixels when integrated with a Si back-panel[8].

Given the continued prevalence of scintillator-based indirect detection in modern X-ray inspection systems[9], such as flat-panel detectors and CT linear-array detectors, there is a pressing need to develop scintillator-based energy-resolving detectors. However, the availability of energy-resolving scintillator detectors with multiple energy channels remains rather limited. For example, dual-energy CT imaging often requires the rapid switching of two X-ray sources[4a, 10], which hampers dynamic imaging capabilities. While stacked dual-layer or multilayer scintillators show promise for multispectral dynamic X-ray imaging[11], improving energy resolution remains a formidable challenge. In this study, an innovative multi-energy X-ray linear-array detector based on a scintillator is presented, which allows for easy acquisition of X-ray spectral information using a visible-light image sensor. Unlike typical energy-integration X-ray imaging where a scintillator is positioned perpendicular to the incident X-rays, our multi-energy scintillator detector, named Side-illuminated X-ray Scintillator (SIXS), is placed parallel to the X-ray beam. This unique arrangement leads to an uneven distribution of scintillation due to the shorter penetration depth of low-energy X-ray photons compared to high-energy X-rays in the material. By leveraging the X-ray attenuation law (Beer's law), the energy spectrum of incident X-rays can be reconstructed by analyzing the intensity distribution of the scintillation.

The emerging $Cs_3Cu_2I_5$ metal halide has been chosen as the basis for our SIXS design. This metal halide exhibits desirable characteristics such as high light yield[12], negligible self-absorption[13], and ease of preparation[14]. In this study, we have both mathematically established and experimentally verified the methodology for discriminating X-ray spectral information using the SIXS detector. Furthermore, a proof-of-concept multi-energy CT imaging system with eight energy channels has been successfully implemented, effectively demonstrating the practical application of our multi-energy linear-array detectors.

## 2. Results and Discussion

A typical energy-integration linear-array X-ray imaging system is portrayed in Figure 1a, where the scintillator and sensor detection plane are positioned perpendicular to the direction of X-rays. In contrast, our side-illuminated X-ray scintillator (SIXS) imaging system, as depicted in Figure 1b, incorporates a scintillator placed parallel to the X-rays. This approach introduces a new dimension of energy-resolving information to the conventional energy-integration X-ray linear-array detector, all without the need for additional hardware components. Figure 1c illustrates the fundamental principle of the energy-resolving mechanism based on SIXS. When a monochrome X-ray beam with an energy of $\mu$ passes through the scintillator, its attenuation adheres to the Beer's law:

$$I(\mu) = I_0(\mu)e^{-\alpha(\mu)*Z} \qquad (1)$$

Where $I_0(\mu)$ is the initial intensity of the X-ray, $\alpha(\mu)$ is the X-ray absorption coefficient, and $Z$ is the penetration depth. In our SIXS setup, the scintillator is divided into multiple regions labeled as *1,2, ..., n*, each with an equal width of *d*. Since the X-ray is incident from the side of the scintillator, the penetration depth when it reaches the nth region is $n*d$. Therefore, the intensity of a monochromatic X-ray with energy $\mu$ after passing through the nth region can be expressed as:

$$I(\mu, n) = I_0(\mu)e^{-\alpha(\mu)*n*d} \qquad (2)$$

The attenuation of the monochromatic X-ray in region *n* can be expressed as:

$$\Delta I(\mu, n) = I(\mu, n-1) - I(\mu, n) = I_0(\mu)[e^{-\alpha(\mu)*(n-1)*d} - e^{-\alpha(\mu)*n*d}] \qquad (3)$$

If the secondary photon and electron-hole pair migration are ignored, the internal scintillation efficiency for an X-ray photon with energy $\mu$ is assumed to be a constant, denoted as $f_\mu$[15]. Considering that X-ray tubes emit a broad range of X-ray photons, the continuous X-ray energy spectrum was divided into *n* energy channels ($u_1$~$u_n$). The total radioluminescence (RL) intensity of the nth region can be calculated as follows:

$$R_n = \sum_{\mu_1}^{\mu_n} \Delta I(\mu, n)f(\mu) = \sum_{\mu_1}^{\mu_n} I_0(\mu)[e^{-\alpha(\mu)*(n-1)*d} - e^{-\alpha(\mu)*n*d}]f(\mu) \qquad (4)$$

Define the coefficient $K(\mu, n)$ as shown in the equation below:

$$K(\mu,n)=[e^{-\alpha(\mu)*(n-1)*d} - e^{-\alpha(\mu)*n*d}]f(\mu) \qquad (5)$$

So, equation (4) can be expressed as:

$$R_n = \sum_{\mu_1}^{\mu_n} I_0(\mu) K(\mu,n) \qquad (6)$$

It can be written as a matrix equation as follows:

$$\begin{pmatrix} R_1 \\ R_2 \\ \vdots \\ R_n \end{pmatrix} = \begin{pmatrix} K(\mu_1,1) & K(\mu_2,1) & \ldots & K(\mu_n,1) \\ K(\mu_1,2) & K(\mu_2,2) & \ldots & K(\mu_n,2) \\ \vdots & \vdots & \ddots & \vdots \\ K(\mu_1,n) & K(\mu_2,n) & \ldots & K(\mu_n,n) \end{pmatrix} \begin{pmatrix} I_0(\mu_1) \\ I_0(\mu_2) \\ \vdots \\ I_0(\mu_n) \end{pmatrix} \qquad (7)$$

The equation is expressed in matrix form as:

$$\boldsymbol{R = KI_0} \qquad (8)$$

Therefore, by solving the coefficient matrix **K**, the X-ray energy spectrum $I_0(\mu_1) \sim I_0(\mu_n)$ can be deduced by measuring the RL intensity distribution $R_1 \sim R_n$.

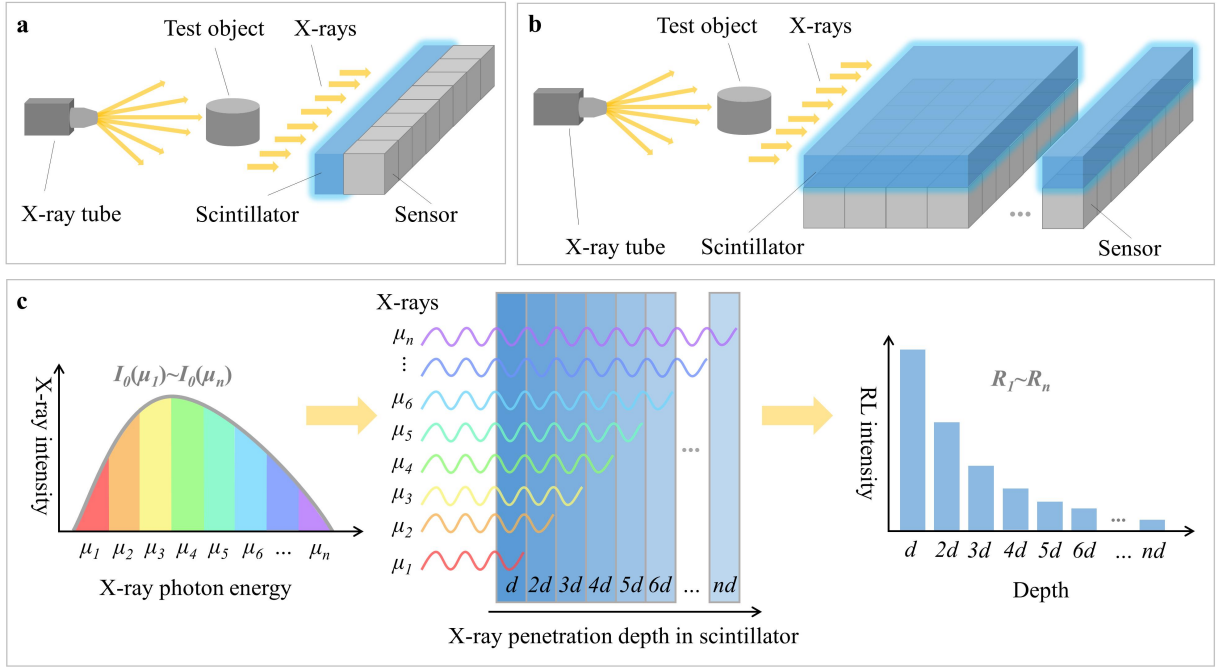

**Figure 1 Schematic of SIXS for multi-energy X-ray detection and imaging**
**(a)** The schematic of the conventional energy-integration linear-array X-ray imaging system. **(b)** The schematic of side-illuminated scintillation for multi-energy X-ray detection and imaging. **(c)** The fundamental principle of X-ray energy discrimination relies on the utilization of a side-illuminated scintillator.

To enable multi-energy X-ray detection and imaging with the SIXS, the selection of a high-performance scintillator is crucial. In this study, the scintillator screen used was fabricated by blending polymethyl methacrylate (PMMA) with $Cs_3Cu_2I_5$ powder. Figure 2a illustrates the RL spectra of the $Cs_3Cu_2I_5$ scintillator screen under the 50 kV tube voltage, exhibiting an emission peak at 452 nm. The inset of Figure 2a provides photographs of the $Cs_3Cu_2I_5$ scintillator screen under daylight and X-ray irradiation, demonstrating a bright and uniform blue emission when excited by X-rays. The scanning electron microscopy (SEM) images reveal the uniform embedding of scintillator particles in the matrix, exhibiting good transparency (Figure S1). Additionally, the scintillation film, protected by PMMA, can be bent and preserved in the air for an extended period (Figure S2). Figure 2b and its inset display the RL intensity of the $Cs_3Cu_2I_5$ scintillator screen at low and high dose rates, respectively, both demonstrating a favorable linear response to the X-ray dose rate. The $Cs_3Cu_2I_5$ scintillator screen has the lowest detection limit (LOD) of 0.12 $\mu Gy_{air}s^{-1}$, ensuring its sensitivity to subtle changes in X-ray intensity and energy spectrum[16]. The monochromatic X-ray (20, 30, 40, 50, 60 keV) attenuation percentages of $Cs_3Cu_2I_5$ at different thicknesses are presented in Figure 2c. The inset shows the absorption coefficient of $Cs_3Cu_2I_5$ as a function of X-ray energy from 0 to 100 keV. The high X-ray absorption coefficient indicates that the side-illuminated scintillation is confined to a small region, which is advantageous for developing more compact detectors[17]. In contrast, commonly used energy-resolving detectors relying on Si cannot achieve this due to Si's limited X-ray absorption capacity (Figure S13)[15]. Furthermore, the scintillator screen shows negligible fluctuations in RL intensity under continuous X-ray illumination at a dose rate of 0.2 $mGy_{air}s^{-1}$, demonstrating good stability (Figure S3). Notably, with continuous irradiation at a high dose rate of 5.0 $mGy_{air}s^{-1}$, the decay in RL intensity for the $Cs_3Cu_2I_5$ scintillator remains impressively within 4.0% when the total dose reaches 9,000 $mGy_{air}$ (Figure S11).

It is important to emphasize that the high spatial resolution of the scintillator in the conventional perpendicular imaging setup plays a crucial role in determining the energy resolution of our SIXS system. In the conventional vertical structure of flat-panel imaging mode, the spatial resolution of the image is compromised due to signal crosstalk among adjacent pixels. In the energy-resolving side-illumination structure, the spectral resolution is constrained by crosstalk from neighboring pixel signals. For instance, when a scintillator material exhibits pronounced self-absorption, photons tend to propagate laterally over a larger distance within the scintillator. This phenomenon results in heightened optical scattering and diminished spatial resolution in traditional vertical imaging mode. Likewise, these factors also contribute to a decrease in spectral resolution in the side-illumination mode. Bear this physical picture in mind, we opted to use the emerging $Cs_3Cu_2I_5$ metal halide as the SIXS.

The Cs$_3$Cu$_2$I$_5$ scintillator exhibits an extremely large Stokes shift (Figure S12) and negligible self-absorption, effectively eliminating cross-talk issues under side-illumination. The spatial resolution of the Cs$_3$Cu$_2$I$_5$ scintillator screen was evaluated in an energy-integration X-ray imaging system (Figure S4). An X-ray image of a standard spatial resolution test board in the partial region was acquired. As demonstrated in Figure S5, the resolution of the scintillator screen was observed to be around 14~16 lp mm$^{-1}$. To further quantify the resolution, the slanted-edge method was employed to obtain the modulation transfer function (MTF) curve by imaging a 1 mm thick aluminum slice with a sharp edge. The spatial resolution (@MTF = 0.2) of the scintillator was determined to be 14.8 lp mm$^{-1}$, consistent with the observation limit (Figure 2d). As a result, the high spatial resolution of the scintillator enables precise energy spectrum measurement. The X-ray images of common objects were acquired using the Cs$_3$Cu$_2$I$_5$ scintillator screen. As shown in Figure 2e, the inner structures of a wireless network card with intricate electronic components were clearly visible. Additionally, an X-ray image of a small headset was successfully obtained (Figure 2f). These experiments demonstrate that the Cs$_3$Cu$_2$I$_5$ scintillator is capable of producing high-quality X-ray images under perpendicular incident X-ray, laying the foundations for multi-energy X-ray detection and imaging by SIXS.

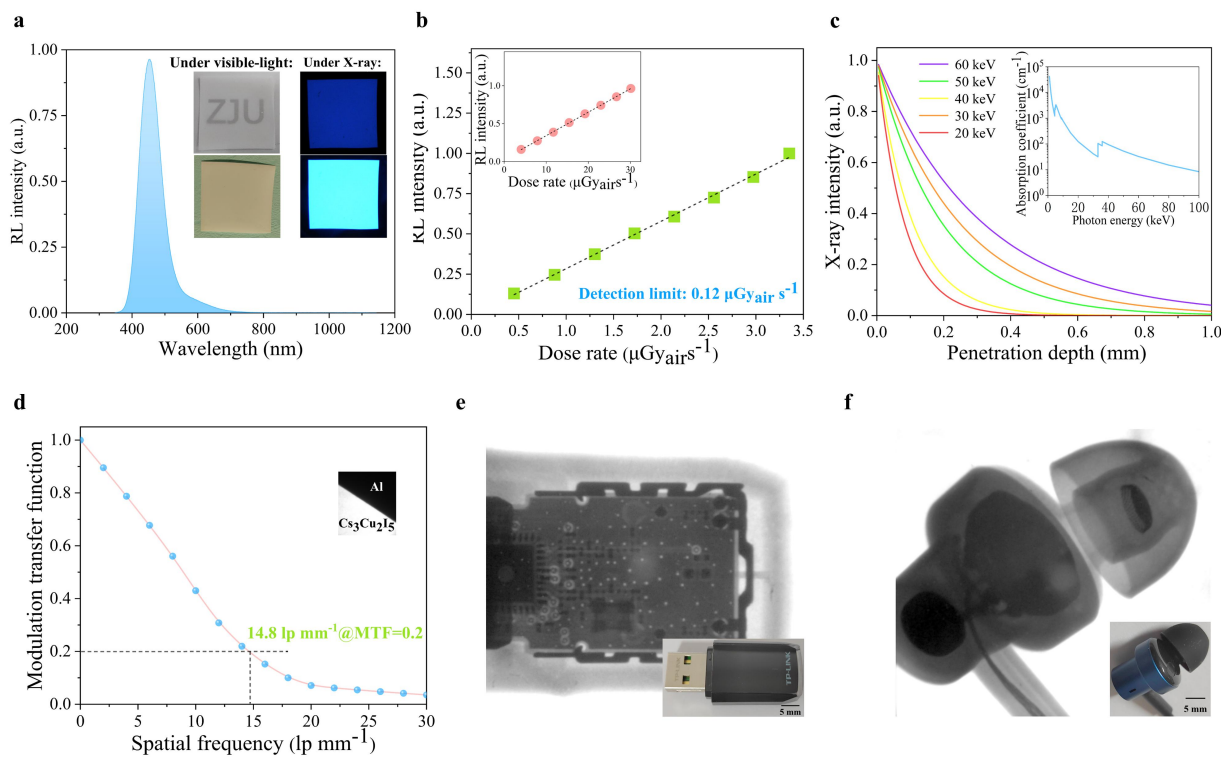

**Figure 2 RL characterizations and X-ray imaging experiments based on Cs$_3$Cu$_2$I$_5$ scintillator.**
**(a)** RL spectra of Cs$_3$Cu$_2$I$_5$. **(b)** The RL linearity of Cs$_3$Cu$_2$I$_5$ under low-dose-rate X-ray excitation. The inset is the linearity of RL intensity at high-dose-rate excitation. **(c)** The X-ray (20, 30, 40, 50, 60 keV) attenuation percentage of $Cs_3Cu_2I_5$ at different thicknesses. The inset is the absorption coefficient of $Cs_3Cu_2I_5$ as a function of X-ray energy. **(d)** The MTF curve of a metallic sharp edge using a $Cs_3Cu_2I_5$ scintillator. **(e)** Photograph of a wireless network card and its X-ray image by $Cs_3Cu_2I_5$ scintillator screen. **(f)** Photograph of a small headset and its X-ray image by $Cs_3Cu_2I_5$ scintillator screen.

In this section, the $Cs_3Cu_2I_5$ scintillator and SIXS model were utilized to achieve multi-energy X-ray detection with eight energy channels. Firstly, ten sets of X-ray energy spectral $I_0$ were generated by modulating the X-ray tube voltages, and the corresponding intensity distribution of side-illuminated scintillation $R$ was measured. The equations $R = KI_0$ were obtained by substituting them into the theoretical model. With the solved coefficient matrix $K$ and the known scintillation intensity distribution $R$, the X-ray energy spectrums $I_0$ can be calculated. Based on the coefficient matrix $K$, four new X-ray spectral were reconstructed, and they exhibited good agreement with the ones measured by an X-ray spectrometer.

The inset of Figure 3a displays the X-ray energy spectrum of the unfiltered tungsten target tube, which exhibits strong characteristic radiation concentrated around 10 keV. However, this concentrated energy distribution is not suitable for demonstrating multi-energy detection, and the soft X-rays are not particularly useful for medical imaging purposes[18]. To address this, an aluminum plate was used as a filter to obtain relatively high-energy bremsstrahlung, which is commonly employed in practical applications. Figure 3a shows the energy spectrum under ten different tube voltages, with X-ray photon energies primarily distributed between 20 and 55 keV. According to the SIXS model, the energy spectrum was divided into eight isometric energy bins, as shown in Figure 3b, which presents two exemplary X-ray spectral captured at 50 and 60 kV tube voltages. The $Cs_3Cu_2I_5$ scintillator was excited by the X-ray of ten different tube voltages, and the resulting RL intensity distributions were recorded. Figure 3c demonstrates that the RL intensity exponentially decays from the illuminated side to the center of the scintillator, in accordance with Beer's law. In this case, the scintillating areas were located within depths of 0~4 mm, and the scintillator was divided into eight signal regions corresponding to eight X-ray energy channels. Ten sets of RL intensity distributions, denoted as $R_1$~$R_8$, were obtained, as shown in Figure 3d. At this point, an overdetermined system of equations with 8×8 unknowns and 8×10 equations was acquired. The target was to find approximate solutions for the overdetermined equations, that is the coefficient matrix $K$. Finally, by multiplying the inverse of the coefficient matrix $K$ with $R_1$~$R_8$, the X-ray energy spectral $I_0(\mu_1)$ ~$I_0(\mu_8)$ can be obtained. As shown in Figure 3e, the average relative error between the calculated $I_0$ and the measured values for all ten X-ray spectral is 2.72%. The illustration of Figure 3e uses the X-ray spectral generated under 50 kV tube voltages as an example to show that the calculated energy spectral is very close to the measured data. Importantly, four new X-ray spectra were predicted by utilizing the determined coefficient matrix ***K***. Figure 3f illustrates that all four calculated energies spectral closely align with the measured values obtained from the X-ray spectrometer (Si drift detector), with an average relative error of less than 5.63%.

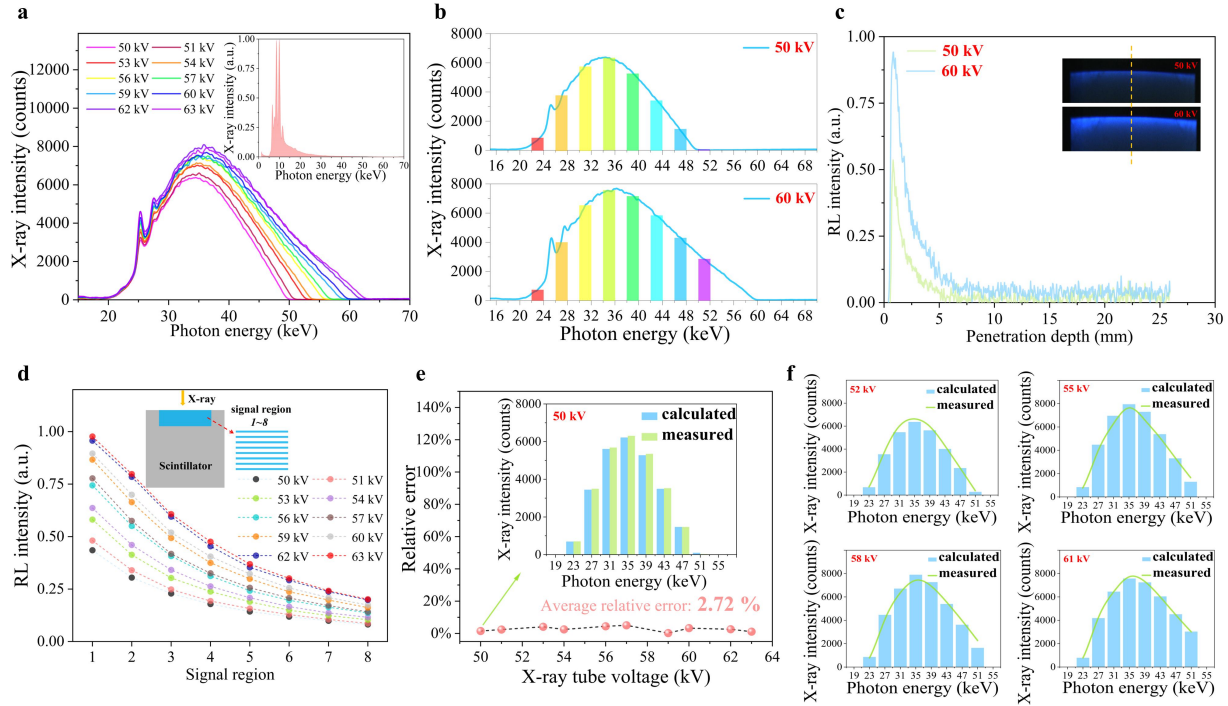

**Figure 3 The X-ray energies are discriminated by solving the scintillation intensity distribution illuminated from the side.**
**(a)** The X-ray energy spectrum of different tube voltages filtered by an aluminum plate. The inset is an unfiltered X-ray energy spectrum at 70 kV tube voltage. **(b)** The X-ray energy spectrum is divided into eight channels according to photon energy, taking 50 and 60 kV tube voltages as examples. **(c)** RL intensity distribution curve of side-illuminated scintillator at 50 and 60 kV tube voltages. **(d)** $R_1$~$R_8$ (mean of RL intensity at signal region *1~8* of the scintillator) under different tube voltages. **(e)** Relative errors between the calculated X-ray energy spectrum and the measured spectrum. The inset shows a comparison between the calculated and actual energy spectrum at 50 kV tube voltage. **(f)** Four new energy spectrums (under tube voltages of 52, 55, 58, and 61 kV) were calculated from side-illuminated scintillation based on the determined coefficient matrix ***K***, they match well with the spectrums measured by the X-ray spectrometer.

In this section, a demonstration of eight channels of multi-energy CT imaging was conducted using SIXS technology. Figure 4a shows the CT imaging system used. A lead (Pb) slit was employed to collimate the X-ray conical beam, ensuring that the scintillator is only excited by X-rays from the desired direction. The $Cs_3Cu_2I_5$ scintillator was positioned horizontally after the Pb slit, and the RL intensity distribution was recorded by a CMOS camera placed beneath the scintillator. An electric rotary table was placed between the slit and the scintillator to rotate the test object. Multi-energy CT imaging was performed on two transverse sections of an electrical plug. Figure 4b illustrates the position of the two transverse sections and their anatomical structure. The side-illuminated scintillation image was recorded every 10°. Images presented in Figure 4c and Figure 4d are the SIXS images of two transverse sections when the scanning angle is 0°. The average RL intensity was calculated for eight signal regions, corresponding to eight X-ray energy channels.

Using the filtering back-projection reconstruction algorithm[19], the multi-energy CT images were reconstructed, as shown in Figure 4e. The internal structures of the two transverse sections in CT images are consistent with the anatomical images. The CT values of six positions on the two transverse sections across different energy channels were analyzed, as depicted in Figure 4f. As the X-ray energy increases, the CT values exhibit a decreasing trend, indicating a decrease in the material's absorption coefficient. Notably, the six testing points can be classified into three patterns based on the slopes of the CT values. Intuitively, materials with higher densities exhibit steeper slopes. Indeed, the positions with the sharpest slopes correspond to copper, while the positions with slightly slower slopes correspond to aluminum. The last two points with the slowest slopes represent plastic. Conventional energy-integration imaging allows easy differentiation between metals and plastics but struggles to distinguish between different types of metals. The study demonstrates the advantage of multi-energy imaging in identifying different substances that have similar image contrasts in energy-integration imaging. The SIXS-based CT imaging does not require additional hardware components compared to energy-integration linear-array CT imaging but adds extra energy-resolving capabilities to the images.

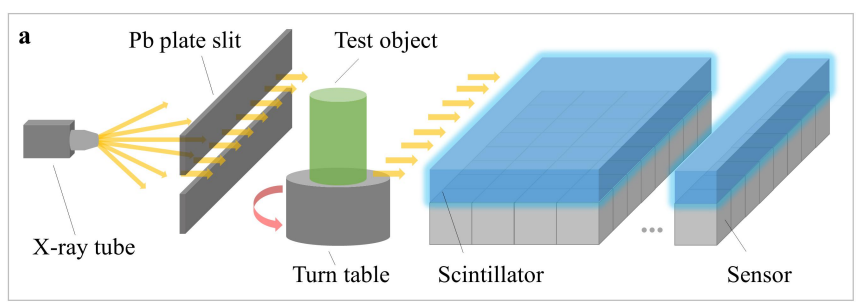

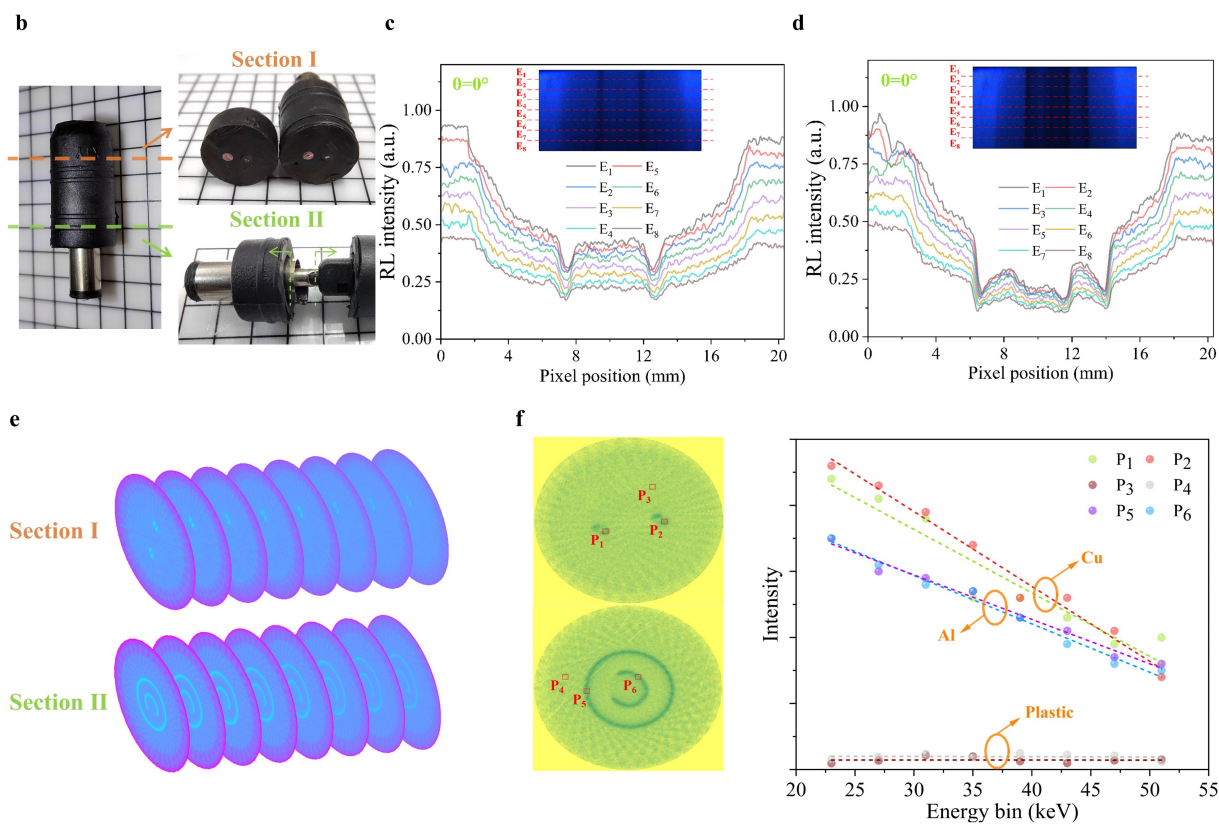

**Figure 4 Multi-energy X-ray CT imaging based on the side-illuminated scintillator.**
**(a)** Schematic of multi-energy CT imaging based on side-illuminated X-ray scintillator system. **(b)** The test object for multi-energy CT imaging is an electrical plug. Two transverse sections of interests are labeled as section I and II. And its anatomical structure images. **(c)** The RL intensity distribution of the side-illuminated scintillator when imaging transverse section I at scanning angle θ=0°. **(d)** The RL intensity distribution of the side-illuminated scintillator when imaging transverse section II at scanning angle θ=0°. **(e)** The multi-energy (8 energy bins) CT images of two transverse sections of the test object. **(f)** CT values at different positions show different dependences on X-ray energies, clearly implying the differences in material densities. The heavier material exhibits a sharper slope.

## 3. Conclusion

In conclusion, we have developed an approach for multi-energy X-ray linear-array detection and CT imaging using the emerging metal halide $Cs_3Cu_2I_5$ scintillator. By analyzing the distribution of scintillation intensity when the scintillator is illuminated from the side, the incident X-ray spectrum can be reconstructed. Compared with conventional energy-integration X-ray linear-array detectors, our proposed method offers an energy-resolving capability without requiring extra hardware components. It is a cost-effective and scalable solution that seamlessly integrates with matured visible-light image sensors. Additionally, it is fully free of the "pile-up" effect and remains suitable for high-flux X-ray applications. A proof-of-concept multi-energy CT imaging system with eight energy channels was successfully demonstrated, achieving a relative error of less than 5.63% between the reconstructed and measured X-ray spectral. These results highlight the enormous potential of our SIXS-based multi-energy detector, particularly in imaging scenarios that employ linear-array detectors. The combination of improved spectral information and compatibility with existing technology makes our approach highly promising for advancing X-ray imaging techniques.

## 4. Experimental Section/Methods

*Chemicals:* Cesium iodide (CsI, 99.9%), copper iodide (CuI, 99.95%), and Methanol (99.8%) were purchased from Aladdin. Methylbenzene (AR) was purchased from Sinopharm Chemical Reagent Co. Ltd. (SCRC). N, N-Dimethylformamide (DMF, 99.9%), and dimethyl sulfoxide (DMSO, 99.9%) were purchased from J&K Scientific. Polymethyl methacrylate was sourced from Kuer Chemical. All reagents and solvents were used without further purification.

*Synthesis of $Cs_3Cu_2I_5$ powder:* The synthesis of $Cs_3Cu_2I_5$ powder was carried out using the anti-solvent method. In this process, 6 mmol of CsI and 4 mmol of CuI were dissolved in 6 ml of DMSO and stirred at 60 °C for 0.5 hours. To prepare, a centrifuge tube was preloaded with 2 mL of methanol as the anti-solvent. Subsequently, 500 μL of the saturated precursor solution was rapidly injected into the tube, resulting in the immediate formation of a substantial amount of white precipitate. The centrifuge tube was then placed in a centrifuge and spun at a speed of 8000 rpm min$^{-1}$ for 5 minutes. Afterward, the supernatant was discarded, and the remaining precipitate was thoroughly washed with methylbenzene. This washing process was repeated more than three times. Finally, the precipitate was dried at 100 °C in a vacuum oven, resulting in the formation of coarse powder.

*Fabrication of $Cs_3Cu_2I_5$ scintillator screen:* First, the coarse powder of $Cs_3Cu_2I_5$, synthesized using the anti-solvent method, was finely ground in an agate mortar. A solution of PMMA-toluene with a concentration of 200 mg mL$^{-1}$ was then stirred at 70 °C for 2 hours. Subsequently, the fine powder of $Cs_3Cu_2I_5$ was added to the mixture at a mass ratio of 1:1 with PMMA. To achieve a uniform dispersion, the mixture required the addition of an appropriate amount of toluene solution since $Cs_3Cu_2I_5$ is insoluble in toluene. Careful stirring and heating were necessary during this process. An improper mass ratio could lead to $Cs_3Cu_2I_5$ aggregation in the film, which would adversely affect the imaging quality. The resulting mixed solution was smoothly and evenly poured into a glass mold, which was then placed in a fume hood at room temperature. After a few hours, the toluene completely evaporated, allowing for the removal of the scintillator film from the mold.

*Material characterization:* The SEM images were captured using a German Zeiss Utral 55 electron microscope. The microscope was equipped with an Oxford X-Max 20 silicon drift detector, enabling high-resolution imaging and analysis of the samples.

*Measurement of radioluminescence (RL) intensity:* A fiber-coupled fluorescence spectrometer (Ocean Optics QE PRO), an X-rays tube, and a quartz mold containing the scintillator powder sample were constructed into an experimental system as shown in Figure S6. The corresponding RL intensity can be calculated by integrating the RL spectrum collected by a spectrometer.

*Energy-integration X-ray imaging system:* The X-ray source utilized in the experiment was a Mini-X2 X-ray tube manufactured by Amptek Inc. It employed tungsten as the target material and had a maximum power output of 10 W. In this setup, X-rays initially passed through the imaging object and were then absorbed by the scintillator screen. To mitigate the adverse effects of direct X-ray radiation on the camera, the optical path was redirected using a reflector. Finally, X-ray images were captured using a high-sensitivity CMOS camera, specifically the Prime 95B model from Photometrics, which was employed to evaluate the imaging performance of the $Cs_3Cu_2I_5$ scintillator screen.

*SIXS system:* In the SIXS system, a color CMOS camera, specifically the ZWO ASI224MC model, was employed due to its compact size and convenience. The camera was used to capture color images of SIXS under X-rays at various tube voltages, as illustrated in Figure S7. These color images were subsequently converted into grayscale images to extract the distribution information of RL intensity on the scintillator. The scintillator surface was divided into signal regions using a specific methodology: starting from the scintillator center of the X-ray illuminated edge, each rectangular area measuring 10*100 pixels was designated as a signal region. Here, the width of the rectangle corresponded to the direction of X-ray penetration. The mean grayscale value of each signal region was then calculated, representing the corresponding $R_n$ value. For the actual measurement of the X-ray energy spectrum under different tube voltages, the X-123 X-ray spectrometer manufactured by Amptek was utilized. The continuous X-ray energy spectrum was divided into eight energy channels. The average

X-ray intensity within each energy channel was used to calculate $I_0(\mu_n)$, as depicted in Figure S8.

*Multi-energy CT imaging system:* During the experiment, X-ray irradiation was conducted at every 10° rotation to capture SIXS images of two transverse sections at different scanning angles, as shown in Figure S9 and Figure S10. For a specific pixel position along the edge of the scintillator, each segment with a width of 10 pixels in the direction of X-ray penetration was designated as a signal region. The mean grayscale value within each signal region was calculated as the projection value for the corresponding energy channel of the X-ray. This process was repeated for SIXS images obtained at each scan angle. Subsequently, the CT reconstructed image of the corresponding energy channel was obtained using the filtering back-projection reconstruction algorithm. By reconstructing the projected values from each energy channel, multi-energy CT images were generated, capturing information from different energy channels.

**Data availability**

The data that support the findings of this study are available from the corresponding author upon reasonable request.

**Acknowledgments**

This research was supported by the Natural Science Foundation of Zhejiang Province of China (LZ23F050005) and the Natural Science Foundation of China (62074136, 52273307).

**Competing Interests Statement**

The authors declare no competing interests.